\documentclass[aps,11pt]{revtex4}
\usepackage[english]{babel}
\usepackage[dvips]{graphicx}
\usepackage{amsmath}
\usepackage{slashed}
\usepackage{braket}
\usepackage{psfrag}

\newcommand {\e}[1]{\mathrm{~#1}}       

\newcommand{\mc}[1]{\mathcal{#1}}

\begin{document}


\vspace{.5cm}

\begin{center}
{\Large \bf On the Dalitz plot analysis of the $B \to K \eta \gamma$ decays}

\vspace{.7cm}

{\large \bf S. Fajfer$^{a,b}$,  T.N. Pham$^{c}$ and N. Ko\v snik$^{a}$\\}
\vspace{0.5cm}
{\it a) J. Stefan Institute, Jamova 39, P. O. Box 3000, 1001 Ljubljana, Slovenia\\}
{\it b)  Department of Physics, University of Ljubljana, Jadranska 19,
1000 Ljubljana, Slovenia\\}
{\it c) Centre de Physique Theorique, CNRS Ecole Polytechnique, 91128 Palaiseau Cedex, France.\\}
\vspace{0.5cm}
{\large \bf ABSTRACT}\\
\end{center}
Recently B-factories have published new results on the $B \to K \eta
\gamma$ decays being inspired by the theoretical suggestion to search
for new physics in $B \to P_1 P_2 \gamma$ decays.  Using heavy meson
chiral perturbation theory we find mechanism which governs the
amplitude in parts of the Dalitz plot where either $K$ or $\eta$
mesons are soft. The dominant contributions in these cases are coming
from the nonresonant decay modes.  We discuss also $B \to K \eta'
\gamma$ Dalitz plot. Our partially integrated rates are 
in agreement with the experimental findings.

\vspace{0.5cm}


\section{Introduction}
In the last decade B physics was one of the tools in the search for
new physics. B-factories made extremely important contributions to these
expectations with the numerous measurements.  An interesting proposal
has been made by the authors of Ref.~\cite{Atwood:2004jj} on the
possible effects of new physics in $B \to P_1 P_2 \gamma$
decays. Namely, in these decays new physics might affect
the right-handed photons. As it is already known in the standard
model~(SM), photon emitted in $b \to s \gamma$ is dominantly
left-handed~\cite{Atwood:1997zr,Grinstein:2004uu}. The photon
polarization can be measured indirectly by means of time-dependent
$CP$ asymmetry of decays to $CP$ eigenstate $f$ plus a photon:
\begin{equation}
  \label{eq:CPV}
  \frac{\Gamma(\bar{B}(t)\to f\gamma) - \Gamma(B(t)\to f\gamma)}{\Gamma(\bar{B}(t)\to f\gamma) + \Gamma(B(t)\to f\gamma)} = S_{f\gamma}\sin(\Delta m t)
  -C_{f\gamma} \cos(\Delta m t).
\end{equation}
Mixing-induced parameters $S_{f\gamma}$ have been studied in radiative
decays of both charged and neutral $B$ decays to $K^*
\gamma$~\cite{Atwood:1997zr}, $B \to P P
\gamma$~\cite{Atwood:2004jj,Grinstein:2005nu}, and also $B \to P V
\gamma$~\cite{Atwood:2007qh}, where $P\,(V)$ is a light
pseudoscalar~(vector) meson.

In this work we focus on $B \to K \eta \gamma$ in kinematical region
with the hard photon (its energy/momentum is of the order $\sim m_b$)
and one soft pseudoscalar (its energy/momentum is of the order $\sim
\Lambda_{QCD}$). Such kinematical conditions call for using chiral
symmetry for soft pseudoscalar and heavy quark effective theory (HQET)
combined with the large energy effective theory (LEET) for heavy meson
and energetic pseudoscalar.  We predict differential decay widths in
these regions.  This channel has been already seen in Belle and BaBar
experiments~\cite{Nishida:2004fk,Aubert:2006vs,Aubert:2008js}, with
the branching fractions~\cite{Aubert:2008js}
\begin{subequations}
\label{bfs}
\begin{align}
  \mc{B} (B^0 \to K^0 \eta \gamma)\times 10^6 &= 7.1^{+2.1}_{-2.0}\pm 0.4,\label{B0toK0etagamma}\\
  \mc{B} (B^+ \to K^+ \eta \gamma)\times 10^6 &= 7.7\pm 1.0 \pm 0.4
\end{align}
\end{subequations}
Quoted errors are statistical and systematic, respectively. However,
$CP$ asymmetries are still consistent with zero although experimental
resolution is about an order of magnitude above the SM expectation.
For three-body decay $\bar{B}^0 \to K_S \pi^0 \gamma$ the authors of
Ref.~\cite{Grinstein:2005nu} used Soft Collinear Effective
Theory~(SCET) in the region with soft pion. They used the Breit-Wigner
ansatz for the resonant channel via intermediate $K^*\gamma$ and
concluded that right-handed photons are mainly due to the resonance
and related interference effects. 

Looking into PDG~\cite{pdg} one finds only two strange resonances with
spin 2 and 3 which potentially contribute to the $\bar{B}^0 \to
\bar{K}^0 \eta \gamma$ decays in the low to intermediate $M_{K\eta}$
region. Their effects are small, as for the $K_2^*(1430)$, the product
$\mathrm{Br} (B\to K_2^*(1430) \gamma) \times \mathrm{Br} (K_2^*(1430)
\to K \eta) \sim 10^{-6}$ is one order of magnitude below branching
fractions~(\ref{bfs}). Similar contribution from $K_3^*(1780)$ is
$10^{-8}$. One cannot expect any important contribution coming from
these resonant states.  This has been confirmed by Belle collaboration
in Ref.~\cite{Nishida:2004fk}. On the other hand, spectra of
BaBar~\cite{Aubert:2008js} show some excess of events in the
$1.4\e{GeV} < M_{K\eta} < 1.8\e{GeV}$ region, but due to large error
bars they are still inconclusive. Following this features we do not
include any resonant contributions in our approach.

\section{Framework}
The $b \to s \gamma$ is induced by the $\Delta B = 1$ effective
Hamiltonian~\cite{Buchalla:1995vs} 
\begin{equation}
  \label{eq:effHam}
\mc{H} = -\frac{G_F}{\sqrt{2}}
V^*_{ts} V_{tb} \left[\sum_{i=1}^6 C_i \mc{O}_i +
  C_{7\gamma} \mc{O}_{7\gamma} + C_{8G} \mc{O}_{8G}\right] + \textrm{h.c.}
\end{equation}
The most important contribution in the SM is due to electroweak
penguin operator which couples tensor current between $b$ and $s$
quarks to the electromagnetic tensor
\begin{equation}
  \label{eq:o7}
  \mc{O}_{7\gamma} = \frac{e}{8\pi^2} \left[m_b \bar{s} \sigma_{\mu\nu} (1+\gamma_5) b + m_s  \bar{s} \sigma_{\mu\nu} (1-\gamma_5) b \right] F^{\mu\nu}.
\end{equation}
Final state photons it produces are dominantly left-handed, with
right-handed ones being suppressed by $m_s/m_b$.  Keeping only
$\mc{O}_{7\gamma}$, this suppression is evident also in the
asymmetry~(\ref{eq:CPV}), however, in multibody decays $\mc{O}_2$ can
induce charm-loop mediated $b \to s \gamma g$, with equal rates for
$\gamma_L$ and $\gamma_R$, and lift the suppression to $\sim
10\%$~\cite{Grinstein:2004uu}. For our purpose of calculating decay
width we can neglect the $m_s$ part of~(\ref{eq:o7}) as well as the
$\mc{O}_2$ effects, keeping only left(right)-handed photons from
$b(\bar{b})$ quark.

\begin{figure}[!h]
  \psfrag{Eeta}{$\Large{E_\eta\,(\mathrm{GeV})}$}
  \psfrag{Mketa}{$\Large{M_{K\eta}\,(\mathrm{GeV})}$}
  \centering\includegraphics[width=0.45\textwidth,angle=-90]{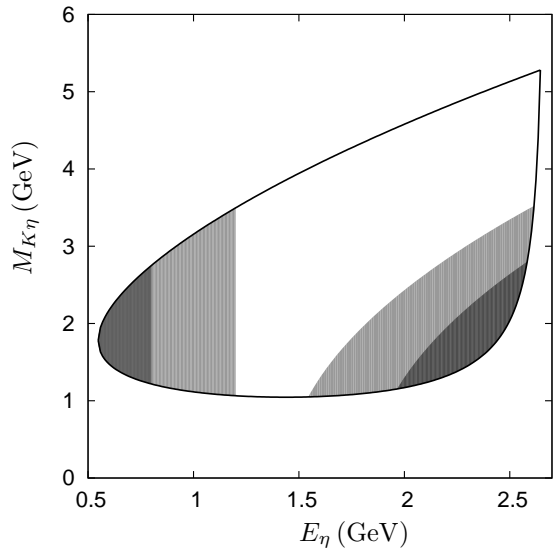}
  \caption{$\bar{B}^0 \to \bar{K}^0 \eta \gamma$ phase space regions
    where soft pseudoscalars have energy below $1.2
    \e{GeV}\,(0.8\e{GeV})$ in the light-gray (gray) region. Left
    corner corresponds to soft $\eta$ and right one to soft $K$.}
\label{fig:ps}
\end{figure}

In decay of $B$ meson to three light particles, there are at least
two energetic final state particles with momentum $\mc{O}(m_b)$.  We
shall study kinematical region of soft $\eta$ and energetic $K$, or
the other way around, while the photon will always be hard, as shown
on Fig.~\ref{fig:ps}, where $E_\eta$ and $K \eta$ invariant mass are
used as kinematical variables.
\begin{figure}[!h]
 \centering 
 \begin{tabular}{cc} \includegraphics[width=0.45\textwidth]{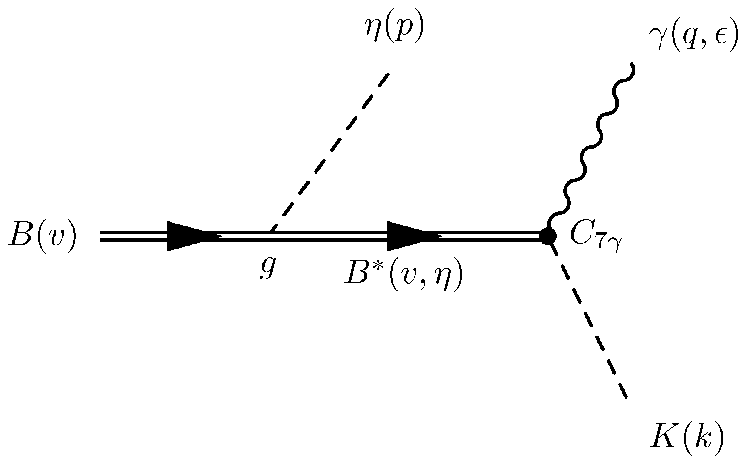} &
   \includegraphics[width=0.45\textwidth]{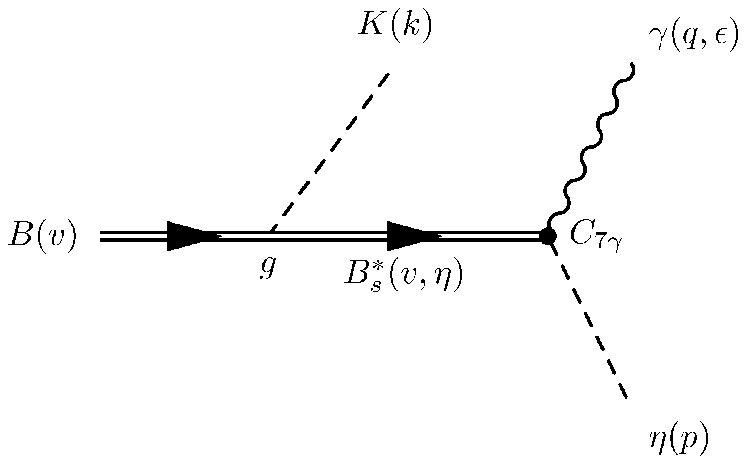}
\end{tabular}
\caption{On the left, the leading contribution in the region of soft
  $\eta$. On the right, $K$ is soft. They govern the decay amplitude
  in the left and right region in Fig.~\ref{fig:ps}, respectively.}
\label{fig:LOdiagram}
\end{figure}

Feynman graphs in the leading order in
$\frac{p_\mathrm{soft}}{\Lambda_\chi},\frac{\Lambda_{QCD}}{m_b}$ are
shown in Fig.~\ref{fig:LOdiagram}, where heavy meson emits a soft
pseudoscalar and is excited to a vector state that decays due to
$\mc{O}_{7\gamma}$ to energetic photon and meson. We stress that those
two diagrams are for two different final states, i.e. with different
momenta, and their sum has no physical meaning. Each corresponds to
precisely defined kinematical region where light meson, attached to
heavy line, has low momentum. This is in contrast to the analogous
decay $\bar{B} \to \bar{K}^0 \pi^0 \gamma$~\cite{Grinstein:2005nu},
where one cannot apply effective description in the soft $K$ region,
due to lack of $s\bar{s}$ component in $\pi^0$.

For strong emission of the soft pseudoscalar off the
heavy-meson line, we utilize the low-energy chiral lagrangian combined
with the heavy-quark symmetry (see~\cite{Casalbuoni:1996pg} and
references therein)
\begin{equation}
  \label{eq:BBstarPi}
  \mc{L}_\mathrm{strong} = i g \,\mathrm{Tr}\left[H_a(v) \mc{A}^\mu_{ab}\gamma_\mu \gamma_5 \bar{H}_b(v) \right].
\end{equation}
The low-energy pion coupling to heavy pseudoscalar and vector has been
calculated on the lattice with unquenched
quarks~\cite{Becirevic:2005zu} and its value is $g=0.5 \pm 0.1$, in
agreement with the value extracted
in~\cite{Fajfer:2006hi}. Contribution of the effective weak vertex
$\mc{O}_{7\gamma}$ in the left graph of Fig.~\ref{fig:LOdiagram} is
\begin{align}
  &\Braket{\bar{K}^0(k) \gamma(q,\epsilon) | \frac{e}{8\pi^2} m_b \bar{s}(0)
    \sigma^{\mu\nu}F_{\mu\nu}(0) (1+\gamma_5) b(0) | B^*(p,\eta)}\\
  &=\Braket{\gamma(q,\epsilon) | 2 \partial_\mu A_\nu | 0} \times
  \Braket{\bar{K}^0(k)| \frac{e}{8\pi^2} m_b \bar{s} \sigma^{\mu\nu}
    (1+\gamma_5)
    b | B^*(p,\eta)} \nonumber\\
  &= \frac{i e m_b}{4\pi^2} q_\mu \epsilon^*_\nu \Braket{\bar{K}^0(k)| \bar{s}
    \sigma^{\mu\nu} (1+\gamma_5) b | B^*(p,\eta)}\nonumber.
\end{align}
For soft $\bar{K}^0$ (right graph of Fig.~\ref{fig:LOdiagram}), the
above manipulations are performed on flavor rotated states
$(B_s^*,\eta) \leftrightarrow (B^*,\bar{K}^0)$. Virtuality of intermediate
$B^*$ is zero up to $1/m_b$ corrections, so use of the heavy-quark
spin-symmetry is justified up to hard spectator
effects~\cite{Beneke:2000wa}. In this picture, we assume heavy-quark
interacts with light degrees of freedom solely through soft gluon
exchanges and thus we use only upper-components field $h_v$ for the
$b$-quark. This is similar to approaches
in~\cite{Charles:1998dr,Beneke:2000wa}. In the following, we are going
to relate the $B^* \to \bar{K}^0$ tensor form-factors to the vector ones of $B
\to \bar{K}^0$. Standard form factors are
\begin{subequations}
\label{eq:FF}
\begin{align}
  \Braket{\bar{K}^0(k) | \bar{s} q_\mu \sigma^{\mu\nu} | B^*(p_B,\eta)} =& 2
  T_1^{BK}(q^2) \epsilon^{\nu\mu\rho\sigma}p_{B,\mu} k_\rho
  \eta_\sigma,\label{eq:tensorFF}\\
  \Braket{\bar{K}^0(k) | \bar{s} q_\mu \sigma^{\mu\nu} \gamma_5 b|
    B^*(p_B,\eta)} =& i T_2^{BK}(q^2) \left[(M^2-m_K^2) \eta - \eta
    \cdot q
    (p_B+k)\right]^\nu \nonumber\\
  &+i T_3^{BK}(q^2) (\eta \cdot q)\, \left[q -
    \frac{q^2}{M^2-m_K^2} (p_B+k)\right]^\nu,\label{eq:pseudotensorFF}\\
  \Braket{\bar{K}^0(k) | \bar{s} \gamma^\nu b| B(p_B)} =& f_+^{BK}(q^2)
  \left[p_B + k - \frac{M^2-m_K^2}{q^2} q\right]^\nu + f_0^{BK}(q^2)
  \frac{M^2-m_K^2}{q^2} q^\nu,
\end{align}
\end{subequations}
where $M$ and $m_K$ are the $B$ and $K$ meson masses, respectively,
and $q = p_B-k$. Now we can use underlying heavy quark and large
energy symmetries to constrain the number of independent form
factors. Following~\cite{Beneke:2000wa}, we express the matrix element
between $B$ and energetic $\bar{K}^0$ as Dirac-trace of their wave functions
\begin{equation}
\label{beneke}
  \Braket{\bar{K}^0(E n_-) | \bar{s}_n \Gamma h_v | B^{(*)} (M v)} = \mathrm{Tr}
\left[A(E) \overline{\mc{M}}_K \Gamma \mc{M}_B\right].
\end{equation}
$E = \frac{M^2 + m_K^2-q^2}{2 M}$ is energy of the $K$ and $n_-$ is
four vector almost parallel to $K$ momentum
\begin{equation}
  k = E n_- + k',\qquad  n_-^2 = 0.
\end{equation}
Residual momentum $k'$ is of the order $\Lambda_{QCD}/E$. $s_n$ is the
effective large-energy field of the $s$ quark
\begin{equation}
  s_n(x) = e^{i E n_-\cdot x} \frac{\slashed{n}_- \slashed{n}_+}{4} s(x),
\end{equation}
and $n_+ = 2 v - n_-$. Long distance physics is parameterized by
function $A(E)$, which does not depend on $\Gamma$, since Hamiltonians
of HQET and LEET commute with quark spin operators. The most general
parameterization of $A(E)$ is then in terms of the four
energy-dependent functions~\cite{Beneke:2000wa}:
\begin{equation}
  A(E) = a_1(E) + a_2(E) \slashed{v} + a_3(E) \slashed{n}_- 
  + a_4(E) \slashed{n}_-\slashed{v}.
\end{equation}
For wavefunctions of mesons, we use
\begin{equation}
  \overline{\mc{M}}_K = -\gamma_5 \frac{\slashed{n}_- \slashed{n}_+}{4},
\qquad
\mc{M}_B = \frac{1+\slashed{v}}{2}\left\{\begin{array}{lcl}
 \slashed{\eta} & ; & B = B^*(M v,\eta)\\
 (-\gamma_5) & ; & B = B(M v)
\end{array}\right. .
\end{equation}
Evaluating the traces on the right-hand side of (\ref{beneke}), one
can connect form factors with functions $a_1(E), \ldots, a_4(E)$ and
find at $q^2=0$ the symmetry relation
\begin{equation}
  T_1^{BK}(0) = T_2^{BK}(0) = T_3^{BK}(0) = f_+^{BK}(0).
\end{equation}
Consequently, matrix element of $\mc{O}_{7\gamma}$ for $B^* \to
\bar{K}^0$ transition
\begin{equation}
  \label{eq:weakVertex}
  \Braket{\bar{K}^0(k) | \bar{s} q_\mu \sigma^{\mu\nu} (1+\gamma_5) h_v| B^*(v,\eta)} =
  f_+^{BK}(0) \left[2 M \epsilon^{\nu\mu\rho\sigma} v_\mu k_\rho \eta_\sigma + i M^2 \eta^\nu 
    - i \eta\cdot q (M v+k)^\nu\right]
\end{equation}
is proportional to $f_+^{BK}(0)$, the value of which has been
determined with the light-cone sum rules approach~\cite{Ball:2004ye}
\begin{equation}
  f_+^{BK}(0) = 0.33 \pm 0.04.
\end{equation}
The left diagram in Fig.~\ref{fig:LOdiagram}, valid in the soft $\eta$
region is then
\begin{align}
  \label{eq:softeta}
  \mc{A}_{\eta\textrm{ soft}} = & -i G_F V_{ts}^* V_{tb} C_7(m_b)
  \frac{e m_b}{8\pi^2} f_+^{BK}(0) \frac{g}{f}
  \left(\frac{\cos \theta}{\sqrt{6}}-\frac{\sin\theta}{\sqrt{3}}\right) \nonumber\\
  &\times\frac{(p_\sigma - v\cdot p\, v_\sigma)}{v\cdot p} \left[2 M
    \epsilon^{\nu\lambda\rho\sigma}v_\lambda k_\rho + i M^2
    g^{\sigma\nu} - i(Mv-k)^\sigma (Mv+k)^\nu\right] \epsilon^*_\nu,
\end{align}
where $\theta=-15.4 {}^\circ$ is the $\eta_8-\eta_1$ mixing
angle~\cite{Feldmann:1998vh} and $f=93\e{MeV}$ is the pion decay
constant. Wilson coefficient $C_{7\gamma}$ on energy scale of
$b$-quark is $C_{7\gamma}(\mu=5\e{GeV}) =
-0.30$~\cite{Buchalla:1995vs}. Electromagnetic gauge invariance is
restored in the limit of small $E_\eta$. Right diagram of
Fig.~\ref{fig:LOdiagram} (with soft $\bar{K}^0$) has amplitude of
similar form
\begin{align}
  \label{eq:softK}
  \mc{A}_{K\textrm{ soft}} =  & i G_F V_{ts}^* V_{tb} C_7(m_b)  \frac{e m_b}{8\pi^2}   f_+^{BK}(0) \frac{g}{f} \frac{\sqrt{2}\cos \theta + \sin\theta}{\sqrt{3}} \nonumber\\
  &\times\frac{(k_\sigma - v\cdot k\, v_\sigma)}{v\cdot k} \left[2 M
    \epsilon^{\nu\lambda\rho\sigma}v_\lambda p_\rho + i M^2
    g^{\sigma\nu} - i(Mv-p)^\sigma (Mv+p)^\nu\right] \epsilon^*_\nu,
\end{align}
In comparison to the soft $\eta$ amplitude~(\ref{eq:softeta}), the
soft $K$ amplitude~(\ref{eq:softK}) has interchanged momenta $p
\leftrightarrow k$ and $\eta_8-\eta_1$ mixing factors now originate
from $B_s^* \eta \gamma$ vertex, where we rely on flavor $SU(3)$ symmetry
to estimate form factor $f_+^{B_s\eta}$.

To get amplitude for $\eta'$ in the final state, one only has to modify
$\eta_8-\eta_1$ mixing coefficients in the amplitudes
(\ref{eq:softeta},\ref{eq:softK}) and find for soft $\eta'$
\begin{align}
  \label{eq:softetap}
  \mc{A}'_{\eta'\textrm{ soft}} = & -i G_F V_{ts}^* V_{tb} C_7(m_b)
  \frac{e m_b}{8\pi^2} f_+^{BK}(0) \frac{g}{f}
  \left(\frac{\sin \theta}{\sqrt{6}}+\frac{\cos \theta}{\sqrt{3}}\right) \nonumber\\
  &\times\frac{(p_\sigma - v\cdot p\, v_\sigma)}{v\cdot p} \left[2 M
    \epsilon^{\nu\lambda\rho\sigma}v_\lambda k_\rho + i M^2
    g^{\sigma\nu} - i(Mv-k)^\sigma (Mv+k)^\nu\right] \epsilon^*_\nu.
\end{align}
Momentum of $\eta'$ is denoted by $p$. Amplitude for soft $K$ and
energetic $\eta'$ is
\begin{align}
  \label{eq:softKetap}
  \mc{A}'_{K\textrm{ soft}} =  & -i G_F V_{ts}^* V_{tb} C_7(m_b)  \frac{e m_b}{8\pi^2}   f_+^{BK}(0) \frac{g}{f} \frac{\cos \theta - \sqrt{2}\sin\theta}{\sqrt{3}} \nonumber\\
  &\times\frac{(k_\sigma - v\cdot k\, v_\sigma)}{v\cdot k} \left[2 M
    \epsilon^{\nu\lambda\rho\sigma}v_\lambda p_\rho + i M^2
    g^{\sigma\nu} - i(Mv-p)^\sigma (Mv+p)^\nu\right] \epsilon^*_\nu.
\end{align}
\begin{figure}[!h]
  \psfrag{omega}{$\Large{\omega\,(\mathrm{GeV})}$}
  \psfrag{dGamma}{$\Large{\frac{\mathrm{d}Br}{\mathrm{d}\omega}\,(\mathrm{GeV^{-1}})}$}
  \centering
 \begin{tabular}{cc}
\includegraphics[width=0.35\textwidth,angle=-90]{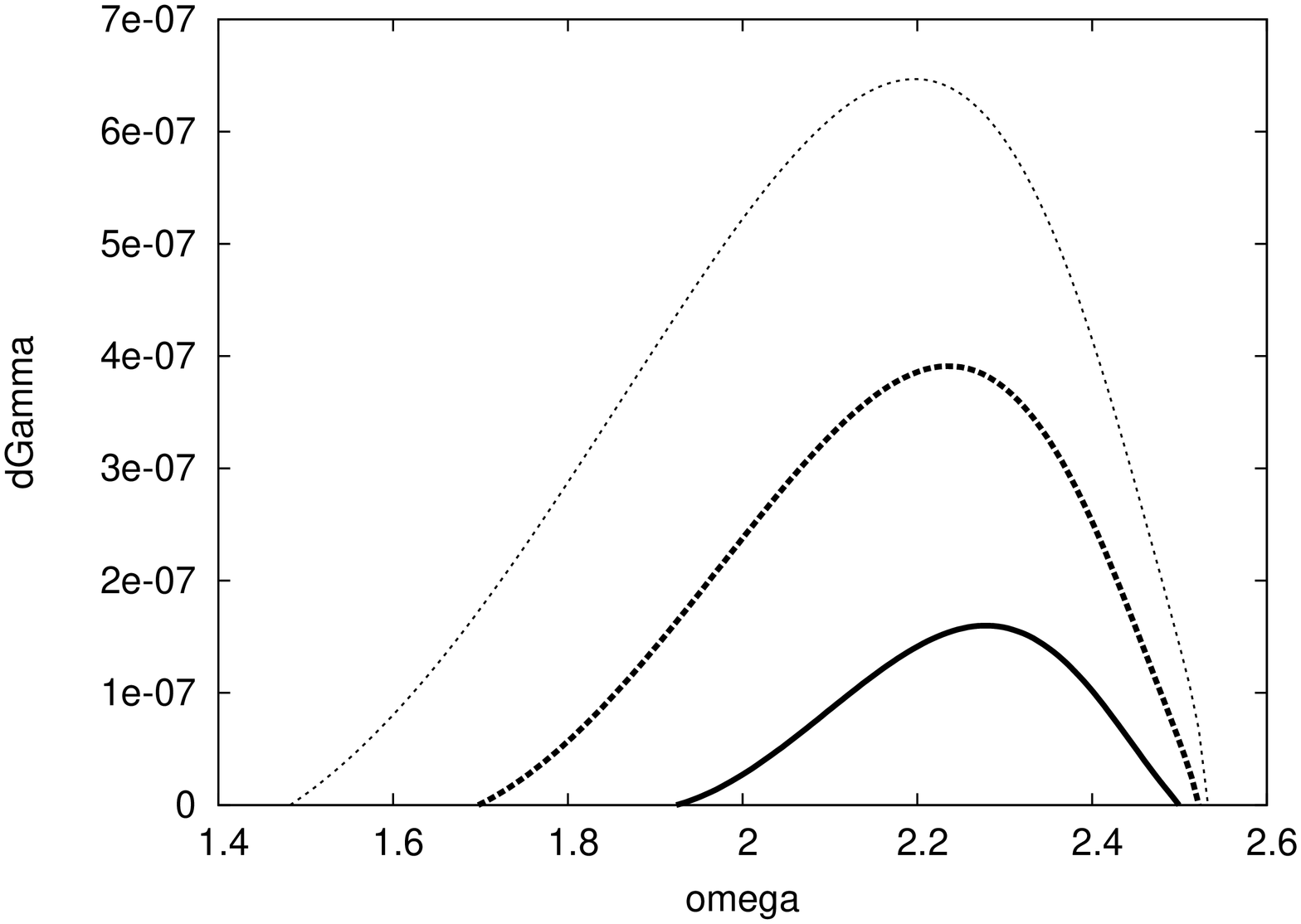} 
&
\includegraphics[width=0.35\textwidth,angle=-90]{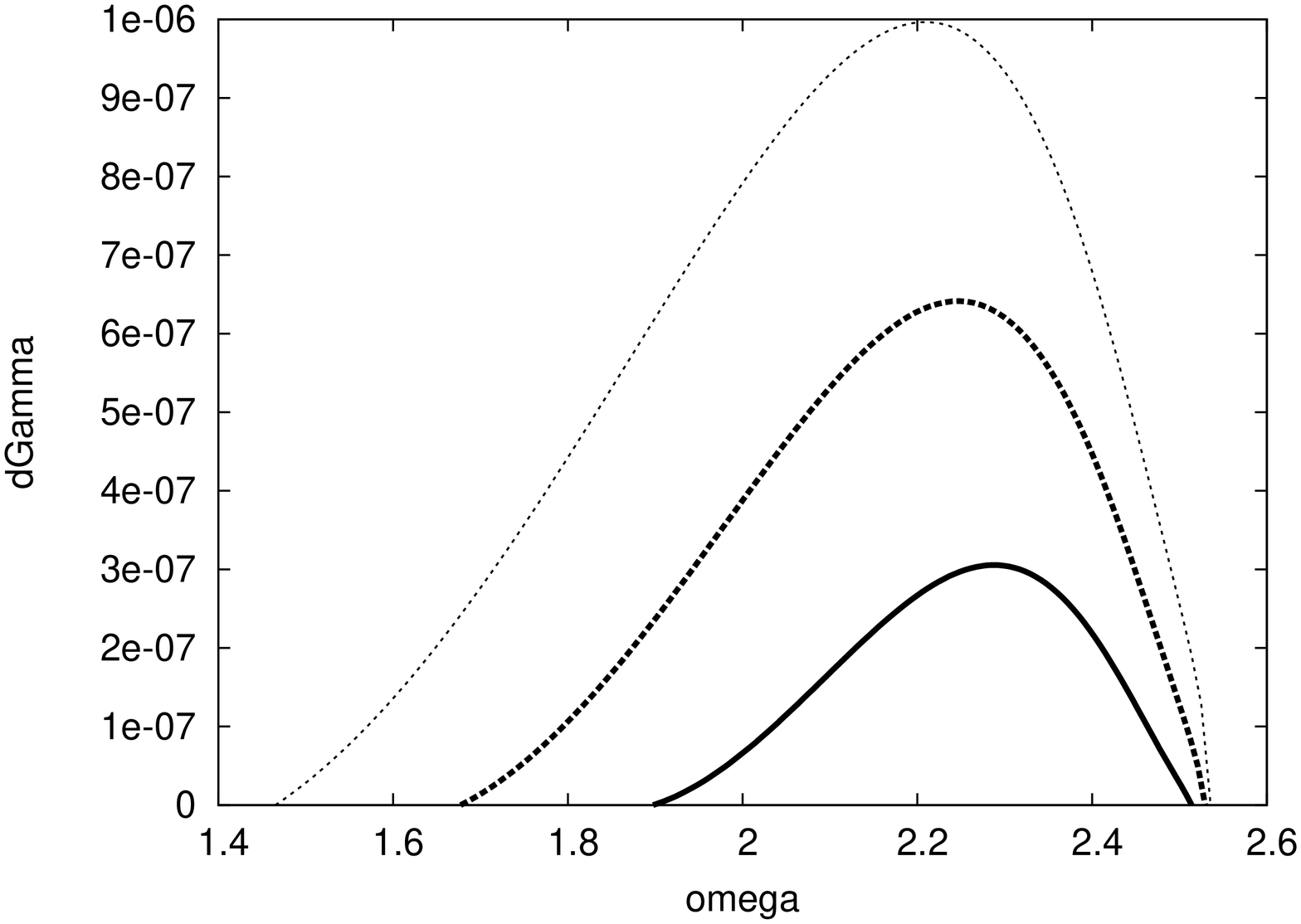}
\end{tabular}
\caption{$\bar{B}^0 \to \bar{K}^0 \eta \gamma$ spectra. Left: Photon
  spectrum in the region of $E_\eta < 0.8\e{GeV}$ (solid thick line),
  $E_\eta < 1.0\e{GeV}$ (dashed thick), and $E_\eta < 1.2\e{GeV}$
  (dotted). Right: same for soft $K$, $E_K < 0.8, 1.0, 1.2\e{GeV}$.}
\label{fig:spectra}
\end{figure}
\begin{figure}[!h]
  \psfrag{omega}{$\Large{\omega\,(\mathrm{GeV})}$}
  \psfrag{dGamma}{$\Large{\frac{\mathrm{d}Br}{\mathrm{d}\omega}\,(\mathrm{GeV^{-1}})}$}
 \centering
 \begin{tabular}{cc} \includegraphics[width=0.35\textwidth,angle=-90]{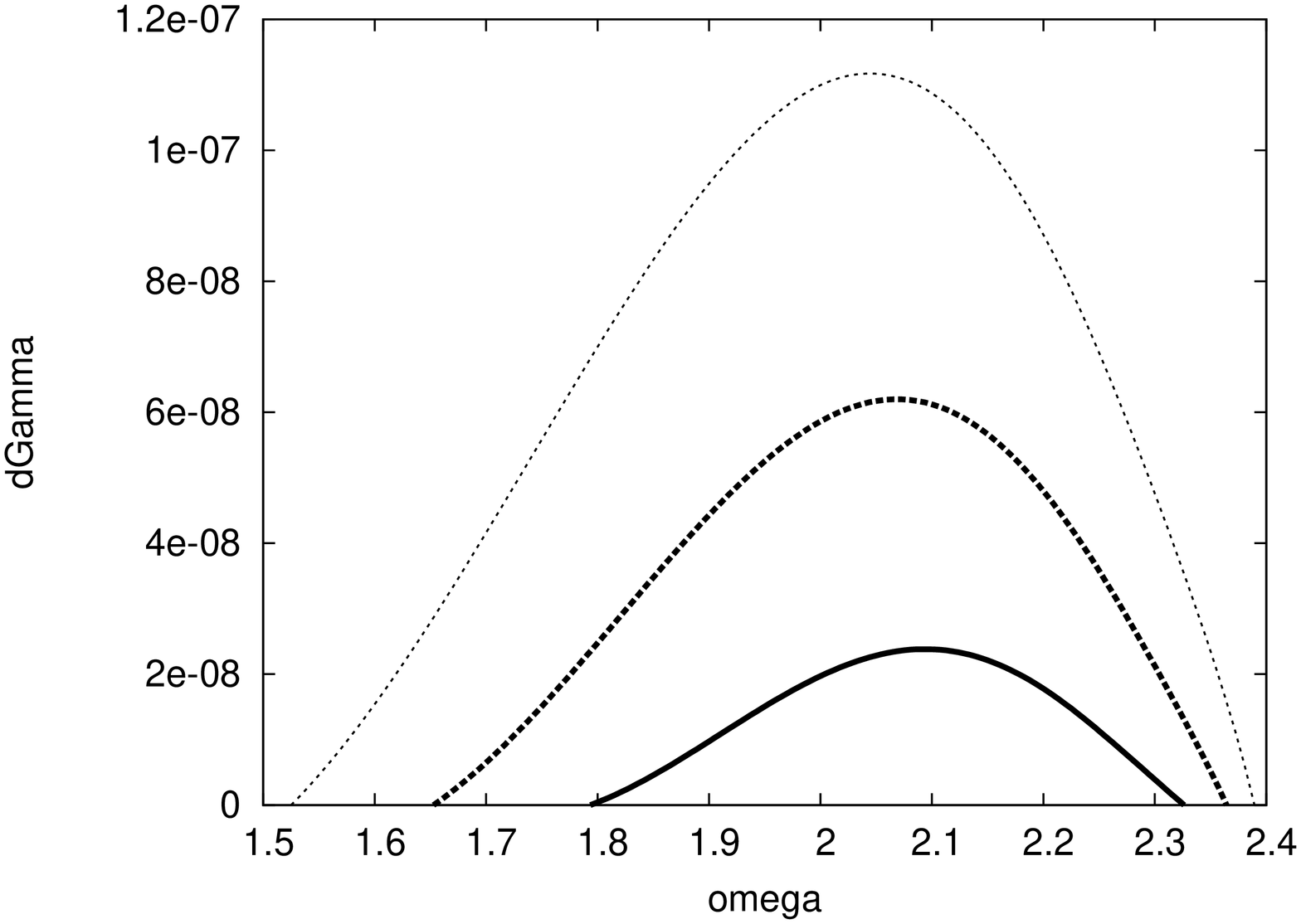} &
   \includegraphics[width=0.35\textwidth,angle=-90]{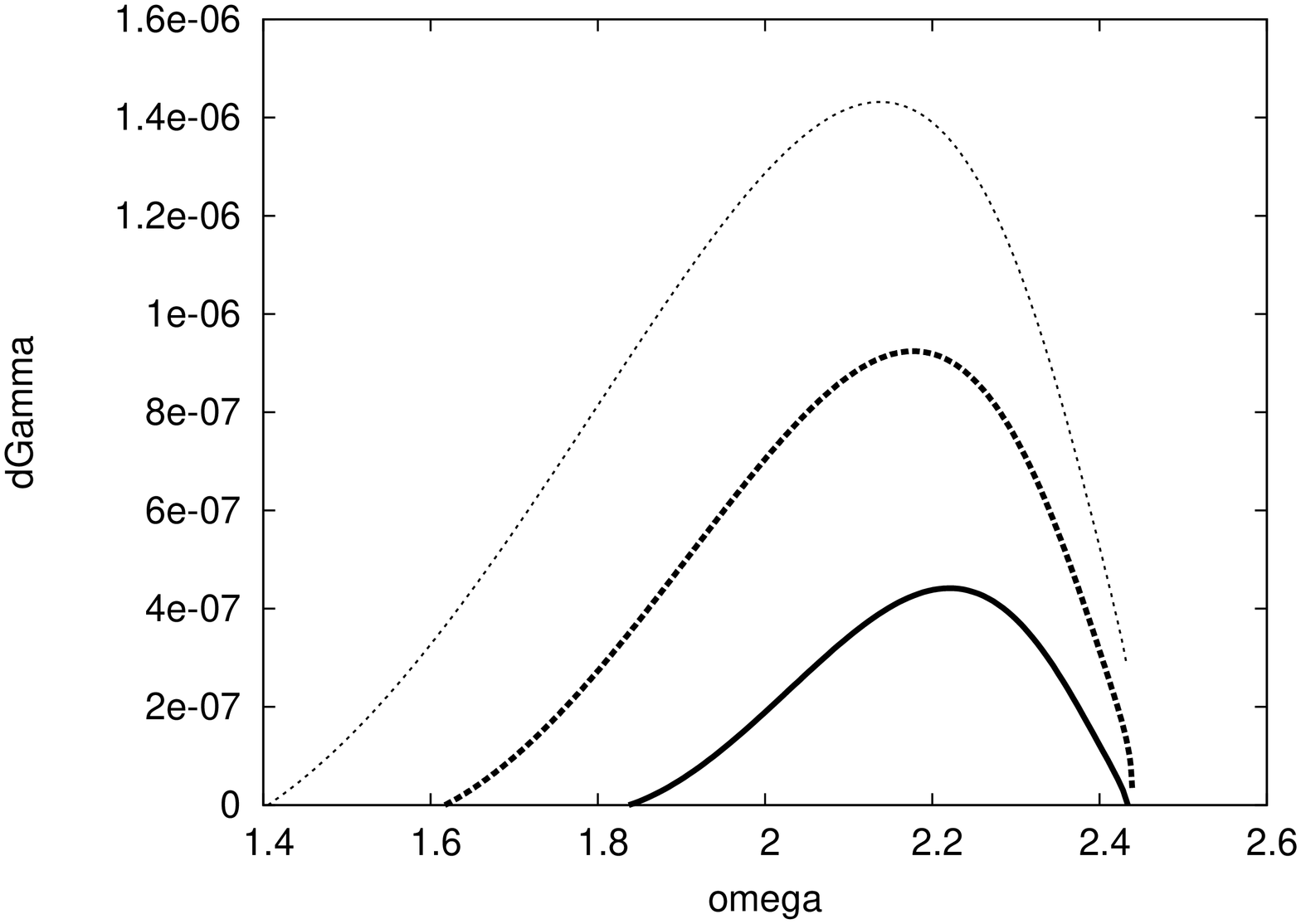}
\end{tabular}
\caption{$\bar{B}^0 \to \bar{K}^0 \eta' \gamma$ spectra. Left: Photon spectrum in the
  region of $E_{\eta'} < 1.1\e{GeV}$ (solid thick line), $E_{\eta'} <
  1.2\e{GeV}$ (dashed thick), and $E_{\eta'} < 1.3\e{GeV}$ (dotted). Right:
  same for soft $K$, $E_K < 0.8, 1.0, 1.2\e{GeV}$.}
\label{fig:spectrap}
\end{figure}

\section{Summary of Results}
We have investigated Dalitz plots of the $B \to \eta (\eta') K \gamma$
decays using the combined heavy meson, large energy, and chiral
lagrangian theories.  The use of this approach is fully justified due
to the fact that in the considered areas of the Dalitz plots, the
kinematical configuration allows simultaneous expansion in soft
momentum and $1/m_b$. Partial branching ratio integrated over both
regions in Fig.~\ref{fig:ps} with upper bound on soft meson energies
set to $1.2\e{GeV}$ accounts for about $10\,\%$ of the $\bar{B}^0 \to
\bar{K}^0 \eta \gamma$ branching ratio~(\ref{B0toK0etagamma}). With
increasing statistics, these two corners of the phase space could be
studied more thoroughly and bring in complementary information on the
scale of $C_{7\gamma}$. On Figures~\ref{fig:spectra} and
\ref{fig:spectrap} we show photon spectra for regions with soft final
state mesons. The model we proposed assumes only nonresonant
production of the $\bar{K}^0 \eta (\eta')$ states. Since $\eta(\eta')$
are isosinglets we do not expect any significant final states effects
and therefore strong phase necessary for the observation of the direct
$CP$ violation is not likely to be generated. Mixing-induced $CP$
violation in $B \to K_S \eta \gamma$, on the other hand, should offer
cleaner environment~(speaking of resonances) to look for right-handed
photons, than the analogous decay $B \to K_S \pi^0 \gamma$.

\section{Acknowledgements}
We thank to Tim Gershon and Jernej Kamenik for their valuable comments
on the first version of this manuscript. This work is supported in
part by the European Commission RTN network, Contract
No. MRTN-CT-2006-035482 (FLAVIAnet). The work of S.F. and N.K. is
supported in part by the Slovenian Research Agency.

\bibliography{references}

\end{document}